\documentclass[doublecol]{epl2} 
\usepackage{amsmath}

\title{The origin of long-range links of air pollution in China}
\shorttitle{The origin of long-range links of air pollution in China} 

\author{Qiuyue Li\inst{1} \and Daqing Li\inst{2} \and Yossi Ashkenazy\inst{3} \and Shlomo Havlin\inst{4}}
\shortauthor{Shlomo Havlin \etal}

\institute{                    
  \inst{1} Department of Physics, Bar-Ilan University - Israel\\
  \inst{2} School of Reliability and Systems Engineering, Beihang University - China\\
  \inst{3} Department of Environmental Physics, Ben-Gurion University of the Negev - Israel\\
  \inst{4} Department of Physics, Bar-Ilan University - Israel\\
}

\abstract{
Weather conditions significantly influence the formation and dispersion of pollution variations. Here we study networks of pollution as well as climate networks and find that pollutants may not only have an impact close to their source but also show a significant correlation with pollutant concentrations thousands of kilometers away. We develop a pollution network model based on cross-correlation between $\mathrm{PM}_{2.5}$ concentration time series in different sites in China to detect stable long-range links during the last ten years. A multi-network analysis of the 500 hPa geopotential height and $\mathrm{PM}_{2.5}$ concentration suggests that long-range correlations in $\mathrm{PM}_{2.5}$ levels are also influenced by synoptic activity.}

\begin{document}

\maketitle

\section{Introduction}
Air pollution has emerged as a critical global challenge due to its multifaceted impacts on human health \cite{davidson2005airborne, kampa2008human, kim2015review}, ecosystems \cite{lovett2009effects, smith2012air}, and climate dynamics \cite{bytnerowicz2007integrated, fiore2012global, von2015chemistry, fiore2015air, francini2024global}. Among various atmospheric pollutants, fine particulate matter with aerodynamic diameters less than or equal to $2.5\mu m$ ($\mathrm{PM}_{2.5}$) has drawn scientific attention due to its critical role in atmospheric haze formation since it was first identified in the early 1990s \cite{chow1995measurement}. Over recent decades, extensive research has examined $\mathrm{PM}_{2.5}$ pollution, particularly focusing on its spatiotemporal characteristics and the meteorological factors influencing its dispersion. Gao et al. \cite{gao2011study} utilized hierarchical cluster analysis to examine air pollution data across 71 Chinese cities, demonstrating that regional air pollution can extend approximately 500 kilometers. Similarly, Hu et al. \cite{hu2014spatial} found that $\mathrm{PM}_{2.5}$ concentrations typically exhibit significant correlations within a radius of about 250 kilometers. Seasonal variations in pollutant concentrations complicate these spatial patterns, as documented by Xie et al \cite{xie2015spatiotemporal}. Beyond seasonal influences, boundary-layer structure and topographic effects have also been shown to modulate pollution accumulation and transport, particularly by influencing vertical mixing and the movement of air masses in complex terrain \cite{sun2020boundary, zhao2019barrier, zhai2019dynamic}. Recent studies have further investigated long-range $\mathrm{PM}_{2.5}$ transport dynamics, highlighting interactions across larger geographical scales and identifying significant transboundary pollution routes, especially within the Beijing-Tianjin-Hebei region \cite{li2019routes}.

Beyond local meteorological conditions, synoptic-scale atmospheric circulation patterns have been increasingly recognized as critical pathways influencing pollutant transport. Stohl et al. \cite{stohl2003backward} employed the Lagrangian particle diffusion model to illustrate intercontinental pollution transport from North America to Europe, emphasizing the widespread distribution of emission sources influencing distant regions. Similarly, Chen et al. \cite{chen2008relationship} and Zhang et al. \cite{zhang2012impact} have highlighted strong correlations between synoptic pressure systems and air quality across northern China. Zhang et al. \cite{zhang2019significant} specifically pointed out that Rossby wave-induced cyclone and anticyclone systems significantly modulate regional air pollution fluctuations through their impacts on atmospheric stability and wind fields. Additionally, recent analyses emphasized the influence of atmospheric teleconnections, such as the Asian Polar Vortex \cite{zhou2020teleconnection} and the East Atlantic/West Russia pattern \cite{li2022winter}, on particulate pollution episodes in northern China. Furthermore, the Arctic Oscillation has been identified as a large-scale atmospheric driver influencing winter $\mathrm{PM}_{2.5}$ variability across China \cite{lu2021impact}.

To better quantify and understand the processes and their impacts on air pollution, advanced statistical methods and analytical frameworks are required. Complex network analysis has recently emerged as an effective framework for understanding such intricate systems. Initially developed to reveal statistical and dynamical properties in diverse systems \cite{barabasi1999emergence, boccaletti2006complex, cohen2010complex, morone2015influence, newman2018networks}, network approaches have proven invaluable for studying climate-related phenomena like El Niño \cite{yamasaki2008climate, tsonis2008topology}, Rossby waves \cite{wang2013dominant}, and extreme weather events \cite{boers2014prediction, agarwal2019network}. Previous applications to air pollution networks have typically focused on local or short-range correlations \cite{hu2014spatial, xie2015spatiotemporal}. However, understanding the stable, long-range correlations in $\mathrm{PM}_{2.5}$ pollution across extensive geographical distances remains limited. Recent studies adopting complex network analyses have started to elucidate the long-range transport routes, clustering behaviors, and spatial spillover features of $\mathrm{PM}_{2.5}$ pollution across China \cite{ying2022complex} and globally \cite{wang2022regional}. These studies have demonstrated that cities within major pollution regions, such as the Beijing–Tianjin–Hebei (BTH) and Yangtze River Delta (YRD), exhibit strong capabilities to export $\mathrm{PM}_{2.5}$ pollution to distant areas, thereby underlining the necessity for regionally coordinated air quality management strategies \cite{ma2021sensitivity}.

To advance our understanding of the complex mechanisms underlying transboundary air pollution, this study employs a cross-correlation network approach to systematically investigate the persistent long-range $\mathrm{PM}_{2.5}$ correlations (pollution network links) observed between sites across China. Specifically, by leveraging a decade-long dataset of high-resolution $\mathrm{PM}_{2.5}$ measurements and geopotential height anomalies at 500 hPa, at a spatial grid in China, we identify stable and significant correlation patterns extending beyond local or regional scales. The present research explicitly quantifies the extent to which large-scale atmospheric circulation patterns, particularly mid-level pressure anomalies, shape and control long-distance pollution linkages. Such insights not only bridge existing knowledge gaps regarding atmospheric transport dynamics but also hold significant implications for improving predictive models and informing coordinated regional air quality management strategies. Ultimately, our findings provide critical scientific information for policy initiatives aimed at effectively mitigating transboundary air pollution.

\section{Methods}

The present study collected and analyzed hourly $\mathrm{PM}_{2.5}$ data from over 400 monitoring stations in China, along with 500 hPa geopotential height and 10m zonal (u) and meridional (v) hourly wind components from the ERA5 reanalysis dataset. The study period spans from 2015 to 2024, covering a ten-year period. To construct a spatially consistent dataset, the hourly $\mathrm{PM}_{2.5}$ concentration for each evenly distributed grid site is calculated as the average of the PM${2.5}$ data from surrounding monitoring stations within a $2.5^\circ \times 2.5^\circ$ range. To ensure data completeness, we first excluded grid sites that had more than 15 consecutive days of missing PM$_{2.5}$ data in any single year over the ten-year period. For the remaining grid sites, missing values were filled using linear interpolation. This resulted in a final selection of 109 grid sites in China that consistently provided data for all ten years. The 500 hPa geopotential height data and 10m wind component data are obtained from the ERA5 global reanalysis dataset ($2.5^\circ \times 2.5^\circ$) provided by the European Centre for Medium-Range Weather Forecasts (ECMWF).

Since meteorological and air pollution time series exhibit seasonal trends, a detrending procedure was applied before computing the Pearson correlation. Specifically, for each hourly time series, we applied a 30-day rolling window to remove seasonal influences. For a given grid site $i$ and its time series $X_i(t)$, the detrended value at time $t$ is computed as:
\begin{equation}
\delta X_i(t) = \frac{X_i(t) - \mu_{30d}(t)}{\sigma_{30d}(t)},
\end{equation}
where $\mu_{30d}(t)$ and $\sigma_{30d}(t)$ represent the mean and standard deviation of $X_i(t)$ of the same hour across the preceding and following 15 days (a total 30-day window). This normalization ensures that the Pearson correlation captures deviations from local seasonal trends rather than absolute variations.

The Pearson correlation function is used to detect connections between meteorological variables of grid sites. The detrended meteorological hourly series of site $i$ and site $j$ during the time period $T$ ($T = 365 \times 24$ hours) are represented by time series $A_i(t)$ and $A_j(t)$ respectively. The cross-correlation function between two time series of grid $i$ and $j$ is defined as
\begin{equation}
C_{i,j}(\tau) = \frac{\langle \delta A_i(t-\tau) \cdot \delta A_j(t) \rangle}{\sqrt{\left\langle \left[\delta A_i(t-\tau)\right]^2 \right\rangle \cdot \left\langle\left[\delta A_j(t)\right]^2 \right\rangle}},
\end{equation}
where $\delta A_i(t)$ is the time series after seasonal detrending. The time delay $\tau \in [-\tau_{max}, \tau_{max}]$ and $\tau_{max} = 720$ hours. We identify the maximum absolute value of $C_{i,j}(\tau)$ as $C_{i,j}^{max}$ and take the corresponding time delay as $\tau^*$. If $\tau^*>0$, the direction of the link is from site $A_i$ to site $A_j$. If $\tau^*<0$, the direction of the link is vice versa. If $\tau^*=0$, it represents an undirected link between them. We define the strength of the correlation as
\begin{equation}
W_{i,j}=\frac{C_{i,j}(\tau^*) - \langle C_{i,j}(\tau)\rangle}{\sigma\left(C_{i,j}(\tau)\right)},
\end{equation}
where $\langle C_{i,j}(\tau) \rangle$ and $\sigma\left(C_{i,j}(\tau)\right)$ is the mean value and standard deviation of the cross-correlation function. The cross-correlation network is constructed based on significant links, for which $W$ and $C_{max}$ are above a threshold chosen according to the shuffled data significance tests. In the present study, we set the 99.9 percentile value of $W$ ($W_{lim}$) for shuffled data as the threshold. The significant links are picked when $W$ is greater than $W_{lim}$.

\section{Results}

Using data from over 400 meteorological observation stations across China, we achieved 109 evenly distributed grid sites and constructed yearly $\mathrm{PM}_{2.5}$-$\mathrm{PM}_{2.5}$ cross-correlation networks over a continuous ten-year period. The results reveal strong significant links between specific grid sites, as demonstrated in Fig 1. Notably, these high correlation links are not limited to adjacent sites but also extend over long distances, with significant links observed between geographically distant locations that exhibit relatively short time delays (Fig. 1a-c). For example, in Fig. 1b, sites A and B, which are separated by approximately 1346 km, exhibit a strong correlation in the pollution in 2021 despite their considerable spatial distance. Further analysis of their cross-correlation function (Fig. 1a) reveals a pronounced peak at $\tau$ = 5 hours, indicating that variations in $\mathrm{PM}_{2.5}$ concentration at one site can be observed at the other site after only a short lag of few hours. 
\begin{figure}
  \centering
  \includegraphics[width=\linewidth]{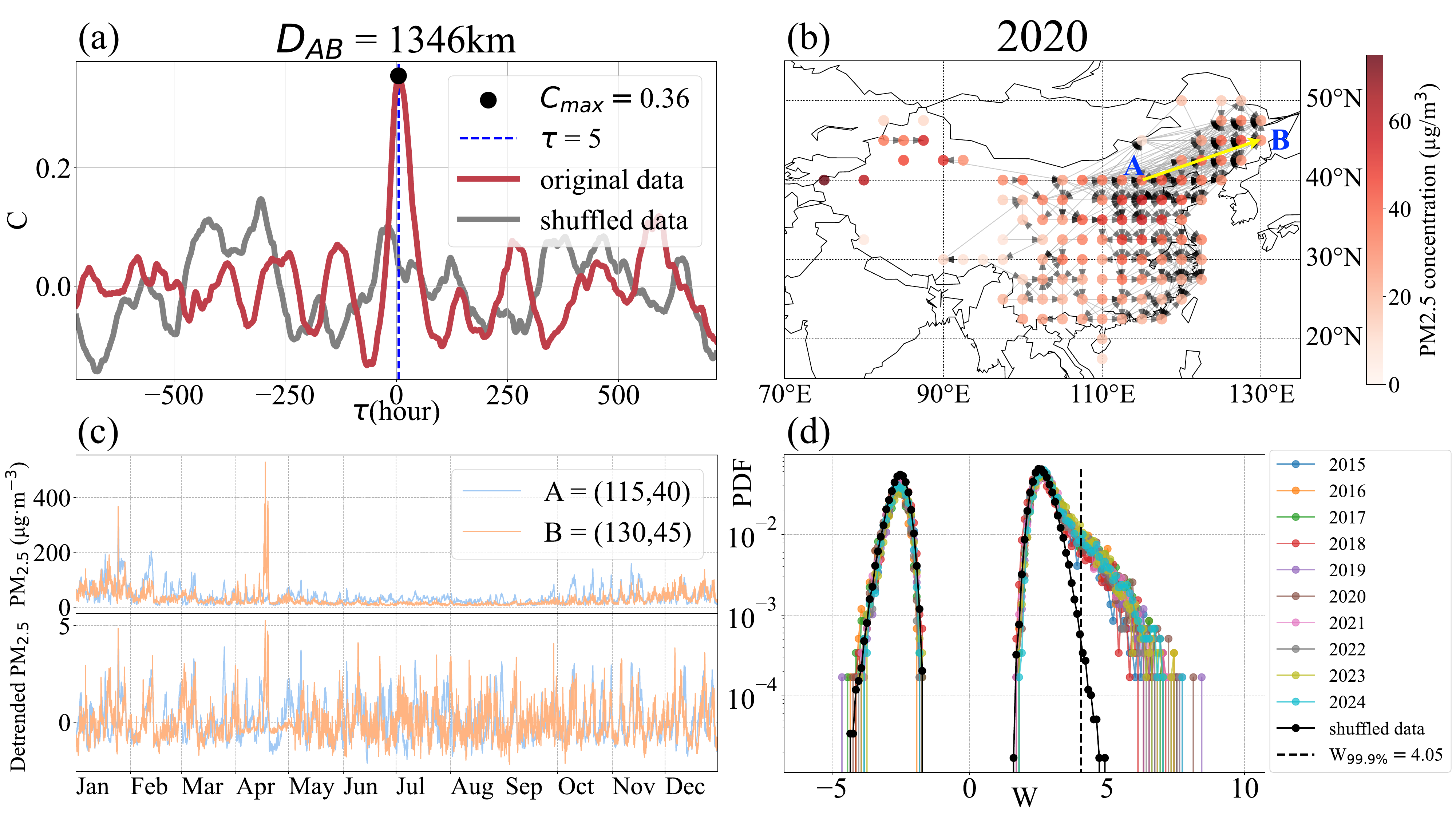}
  \caption{\textbf{PM$_{2.5}$-PM$_{2.5}$ cross-correlation network in China. (a)} Cross-correlation function between two representative grid sites (Site A and Site B shown in panel b) separated by approximately 1346 km. The observed shifted cross-correlations (red curve) exhibits a clear significant positive peak (C = 0.36) at a relatively short time delay ($\tau^{*}$ = 5 hours); this link is significant ($W=4.7$) based on a comparison with the shuffled data (gray curve). \textbf{(b)} Geographic locations of sites A and B, with the significant link between them marked by a yellow arrow. Other significant directed links in the PM$_{2.5}$ network for 2021 are represented by black arrows. The mean PM$_{2.5}$ level of the different sites is indicated by the colorbar. \textbf{(c)} A one-year example illustrating both PM$_{2.5}$ concentration variations (top) and seasonally detrended PM$_{2.5}$ (bottom) at grid sites A and B, clearly visually demonstrating correlated temporal fluctuations consistent with the identified significant link in (a). \textbf{(d)} Distribution of cross-correlation strengths (W) in the PM$_{2.5}$ networks constructed from original (colored curves) and shuffled data (black curve). The dashed black vertical line represents the 99.9th percentile threshold derived from shuffled data, above which links are considered statistically significant.}
  \label{fig.1}
\end{figure}

Additionally, direct inspection of the raw PM$_{2.5}$ concentration time series and normalized time series from both sites (Fig. 1c) supports the co-variability in pollutant levels, consisting with the statistically quantified findings. When compared with shuffled data (Fig. 1d), the original dataset consistently exhibits robust and structured correlation patterns i.e., the probability density function (PDF) of the data is significantly stretched to the right, indicating that the observed links are significant and suggest to reflect an underlying physical mechanism.

Comparing the PM$_{2.5}$ cross-correlation networks constructed from consecutive annual data in different years (2015–2024), we observe a substantial number of stable links recurring consistently throughout this period. To quantify the temporal stability, we calculated the Jaccard Index for significant links across pairs of years, with different spatial distance constraints applied (0–500 km, 500–1000 km, and above 1000 km).
The Jaccard Index is defined as:

\begin{equation} J(Y_i, Y_j) = \frac{|L_i \cap L_j|}{|L_i \cup L_j|} \end{equation}
where $L_i$ and $L_j$ represent the sets of significant links in years $Y_i$ and $Y_j$, respectively. The numerator denotes the number of links that appear in both years, while the denominator represents the total number of unique links observed in either year. The results are shown in Fig. 2a-c.

The relatively high and stable Jaccard Index indicates that certain correlation structures persist over multiple years, highlighting their robustness and suggesting underlying recurring meteorological or atmospheric mechanisms. This structural persistence is particularly notable given the substantial decline in PM$_{2.5}$ concentrations observed during the same time span \cite{wen2024combined, geng2024efficacy}, implying that the inter-site correlation patterns are shaped more by stable atmospheric dynamics than by pollutant levels themselves. Notably, over 700 links appear more than three times over the 10-year period (Fig. 2f), reinforcing the existence of stable correlation structures within China's PM$_{2.5}$ distribution. Additionally, we analyzed the relationship between the average time delay ($\overline{\tau^{*}}$) and spatial distance among grid sites that exhibited persistent significant links over at least three of the ten studied years (Fig. 2e). Intriguingly, the observed time delays associated with many long-distance correlations (especially those exceeding 1000 km) are considerably shorter than the expected travel time of pollutants transported solely by surface wind fields (Fig. 2d, e). This discrepancy strongly suggests the presence of large-scale meteorological drivers or synoptic conditions that simultaneously affect distant grid sites, rather than simple wind-driven pollutant dispersion.

\begin{figure}
  \centering
  \includegraphics[width=\linewidth]{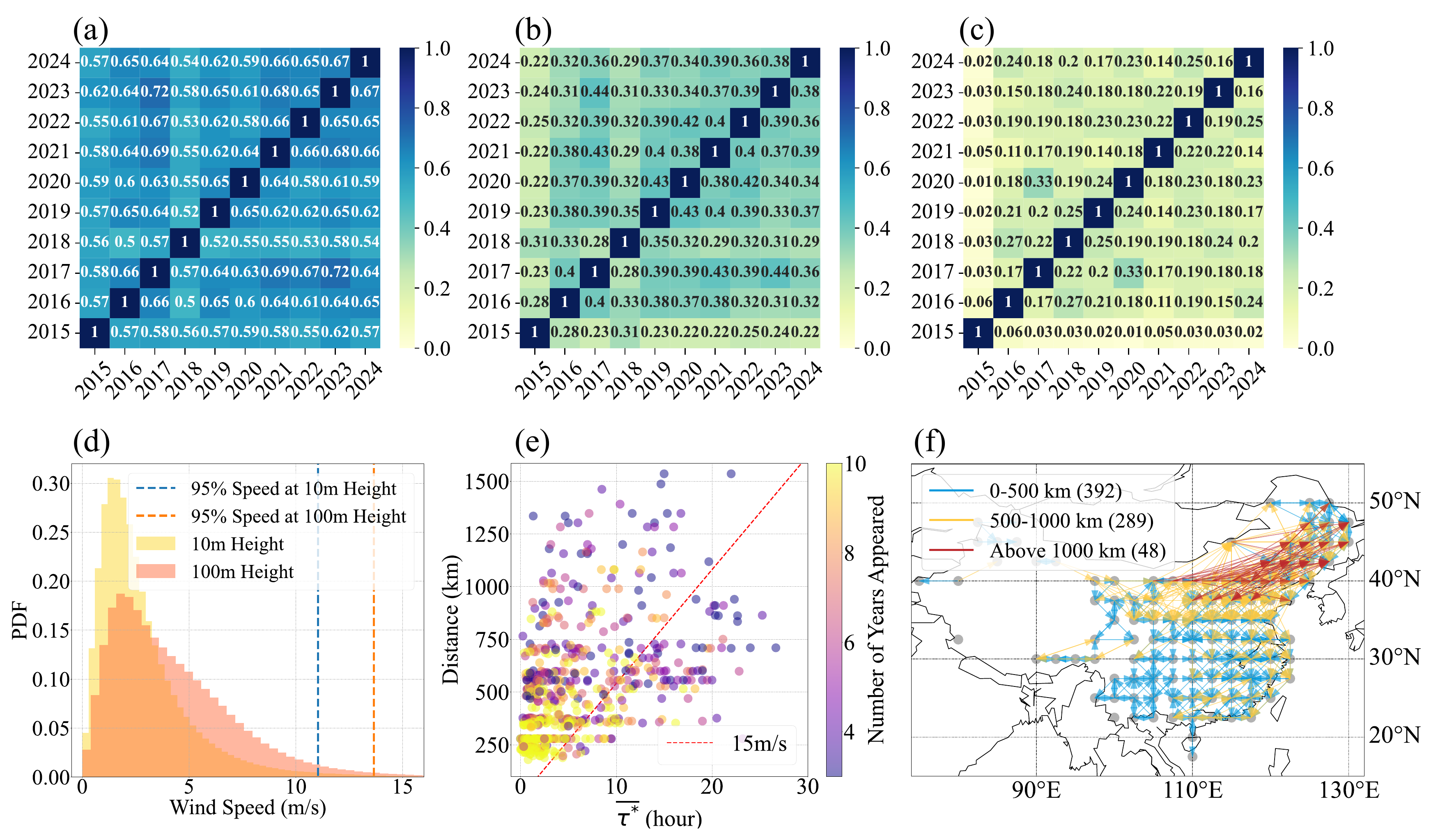}
  \caption{\textbf{Persistence of PM2.5-PM2.5 cross-correlation network in different years in China. (a–c)} Heatmaps showing the Jaccard Index calculated between each pair of annual cross-correlation networks under varying distance thresholds: (a) 0–500 km, (b) 500–1000 km, and (c) distances greater than 1000 km. \textbf{(d)} Wind Speed Distribution at 10m and 100m Heights. The dashed blue line indicates the 95th percentile speed at 10m, while the dashed orange line marks the 95th percentile speed at 100m. \textbf{(e)} Relationship between average time delay ($\overline{\tau^{*}}$) in hours of the links and geographical distance for links repeatedly appearing in at least three out of the ten years studied. The dashed red line represents a hypothetical speed of 15 m/s which far exceeds reasonable surface wind speeds, see (d). Each circle represents a unique link, with the color indicating the frequency of the link's appearance over the years, as denoted by the color bar to the right. \textbf{(f)} Spatial distribution of persistent significant links appearing at least three years throughout the study period. Links shorter than 500 km are shown in light blue, intermediate-range links (500–1000 km) in yellow, and notably long-distance links exceeding 1000 km in red.}
  \label{fig.2}
\end{figure}

A possible mechanism underlying the observed persistent long-range PM$_{2.5}$ links could involve other meteorological variables that simultaneously influence at different times or same time distant PM$_{2.5}$ grid sites, thereby inducing coherent temporal variations in PM$_{2.5}$ and a non-direct link between them. Inspired by prior research highlighting the role of synoptic-scale atmospheric fields\cite{zhang2019significant} on PM$_{2.5}$, we specifically investigate how anomalies in the 500 hPa geopotential height field may affect PM$_{2.5}$ variability across China and the long-range links we observed here, see Fig. 3. As illustrated in Fig. 3d, significant correlations between geopotential height anomalies and PM$_{2.5}$ concentrations form two distinct clusters. One cluster, denoted by red triangles, exhibits positive correlations, suggesting elevated geopotential heights correspond to enhanced PM$_{2.5}$ levels across associated grid sites. To improve visual clarity, grid points in this positive correlation cluster were uniformly shifted northward by 10 degrees latitude. Conversely, the second cluster, indicated by blue triangles, shows predominantly negative correlations, signifying reduced geopotential heights associated with elevated PM$_{2.5}$ concentrations.

\begin{figure}
  \centering
  \includegraphics[width=\linewidth]{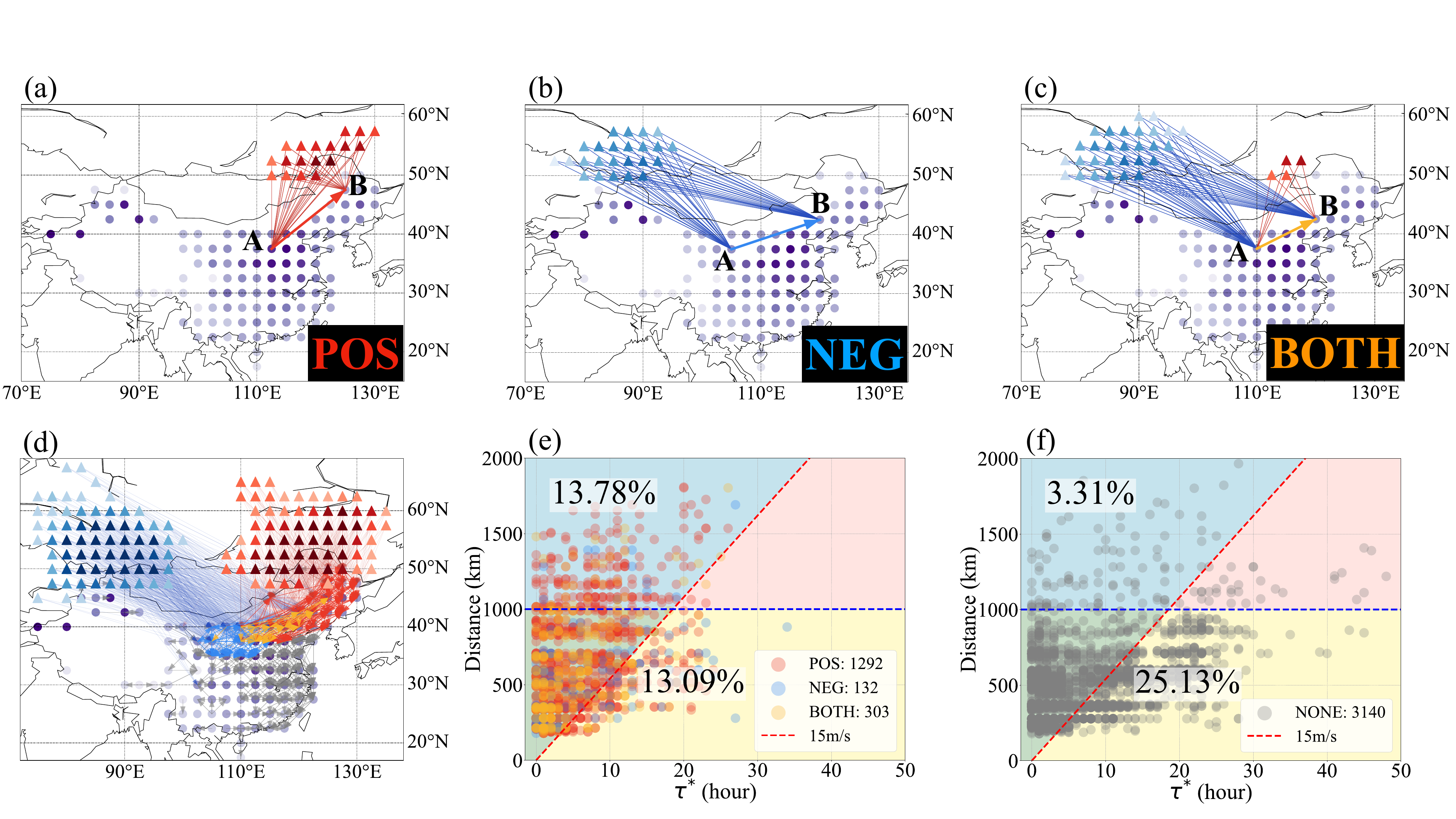}
  \caption{\textbf{Geopotential height (GH, 500 hPa)-PM$_{2.5}$ cross-correlation network. (a-c) Demonstration of three} distinct correlation patterns showing the influence of geopotential height anomalies on two different PM$_{2.5}$ sites A and B: \textbf{(a)} POS-links, exclusively positive geopotential height-PM$_{2.5}$ correlations; \textbf{(b)} NEG-links, exclusively negative correlations; \textbf{(c)} BOTH-links, positive and negative geopotential hight-PM$_{2.5}$ correlations; \textbf{(d)} Spatial distribution of geopotential height-PM$_{2.5}$ correlation patterns in 2017; triangles represent geopotential height anomaly sites (red: positive, blue: negative), and their color shade indicates out-degree (number of influenced PM$_{2.5}$ sites). POS-, NEG-, BOTH-links, and links without geopotential height correlations are colored red, blue, yellow, and gray, respectively. For clearer visualization, all red triangles in panels (a–d) are uniformly shifted 10° northward. \textbf{(e, f)} Scatter plots of time delay ($\tau^*$) versus distance for all significant PM$_{2.5}$ links from 2015–2024. Colored dots represent geopotential height correlated links, categorized as POS (red), NEG (blue), or BOTH (orange), consistent with panels (a–c). Grey dots represent links not associated with any geopotential height sites. The upper-left region (distance above 1000 km, transport speed more than 15m/s) reflects long-distance, fast-propagating links. This region accounts for 13.78\% of geopotential height associated links in (e), but only 3.31\% of non-associated links in (f). The dashed red line indicates an transport speed of 15 m/s.}
  \label{fig.3}
\end{figure}

To systematically understand these relationships, we classify the significant PM$_{2.5}$ links into three distinct correlation patterns according to their associations with geopotential height anomalies (Fig. 3a–c): 
\begin{itemize}
    \item \textbf{POS-link (Fig. 3a):} Both PM$_{2.5}$ sites involved in an AB link exhibit exclusively positive correlations with the same geopotential height sites, suggesting simultaneous responses to large-scale positive geopotential height anomalies.
    
    \item \textbf{NEG-link (Fig. 3b):} Both PM$_{2.5}$ sites A and B share exclusively negative correlations, indicating coherent responses to large-scale negative geopotential height anomalies.
    
    \item \textbf{BOTH-link (Fig. 3c):} PM$_{2.5}$ sites A and B display a mixed response, exhibiting both positive and negative correlations with the same geopotential height sites, highlighting more complex atmospheric interactions.
\end{itemize}

Notably, these geopotential height-driven correlation patterns dominantly appear in northeastern China (Fig. 3d), spatially consistent with the identified stable, long-distance PM$_{2.5}$ correlations (as shown in Fig. 2). Such geographic alignment may suggest that synoptic-scale atmospheric processes are critical drivers behind persistent regional air pollution patterns. 

We further classify all PM$_{2.5}$ links from 2015 to 2024 into two groups: those not associated with any geopotential height anomalies (Fig. 3f, grey dots), and those for which both endpoints are simultaneously correlated with at least one common geopotential height site (Fig. 3e, colored dots). These colored dots follow the POS-, NEG-, and BOTH-link categories defined in Fig. 3a-c. A comparison between Fig. 3e and Fig. 3f highlights an obvious pattern: in the upper-left region, representing long-distance links (above 1000 km) with short time delays (more than 15 m/s transport speed), geopotential height-associated links account for a much larger proportion—13.78\% versus 3.31\%—strongly suggesting that these rapid, long-range PM${2.5}$ connections are more likely driven by large-scale atmospheric circulation linked to mid-tropospheric pressure anomalies. The BOTH-links pattern, in which PM$_{2.5}$ sites show both positive and negative correlations with the same geopotential height anomaly (see e.g. Fig. 3c), likely reflects more complex and dynamic atmospheric processes. One possible mechanism is the influence of frontal systems, which can generate heterogeneous impacts on air quality across adjacent regions. Such transitional meteorological structures may explain why some sites exhibit mixed PM$_{2.5}$ responses (positive and negative) to a shared geopotential driver. Future studies incorporating frontal boundary datasets or regional weather reanalysis could help further test this interpretation.

Our findings are consistent and advances the earlier study \cite{zhang2019significant}, which identified Rossby wave patterns as a major synoptic driver of regional air pollution in China. These waves modulate the large-scale distribution of high- and low-pressure systems, thereby influencing atmospheric stability and vertical motion across wide regions. For instance, the anticyclonic phase of a Rossby wave is typically associated with subsidence and temperature inversions, which suppress vertical mixing and favor pollution accumulation, while the cyclonic phase may enhance vertical motion and pollutant dispersion. Our results extend this mechanism by revealing that the stable, long-range PM$_{2.5}$ correlations observed over multiple years are indeed driven by geopotential height (GH) anomaly clusters, which simultaneously influence distant PM$_{2.5}$ sites. Thus, these persistent remote linkages reflect a higher-order synchronization indirectly governed by common synoptic-scale atmospheric dynamics associated with Rossby waves.

\begin{figure}
  \centering
  \includegraphics[width=\linewidth]{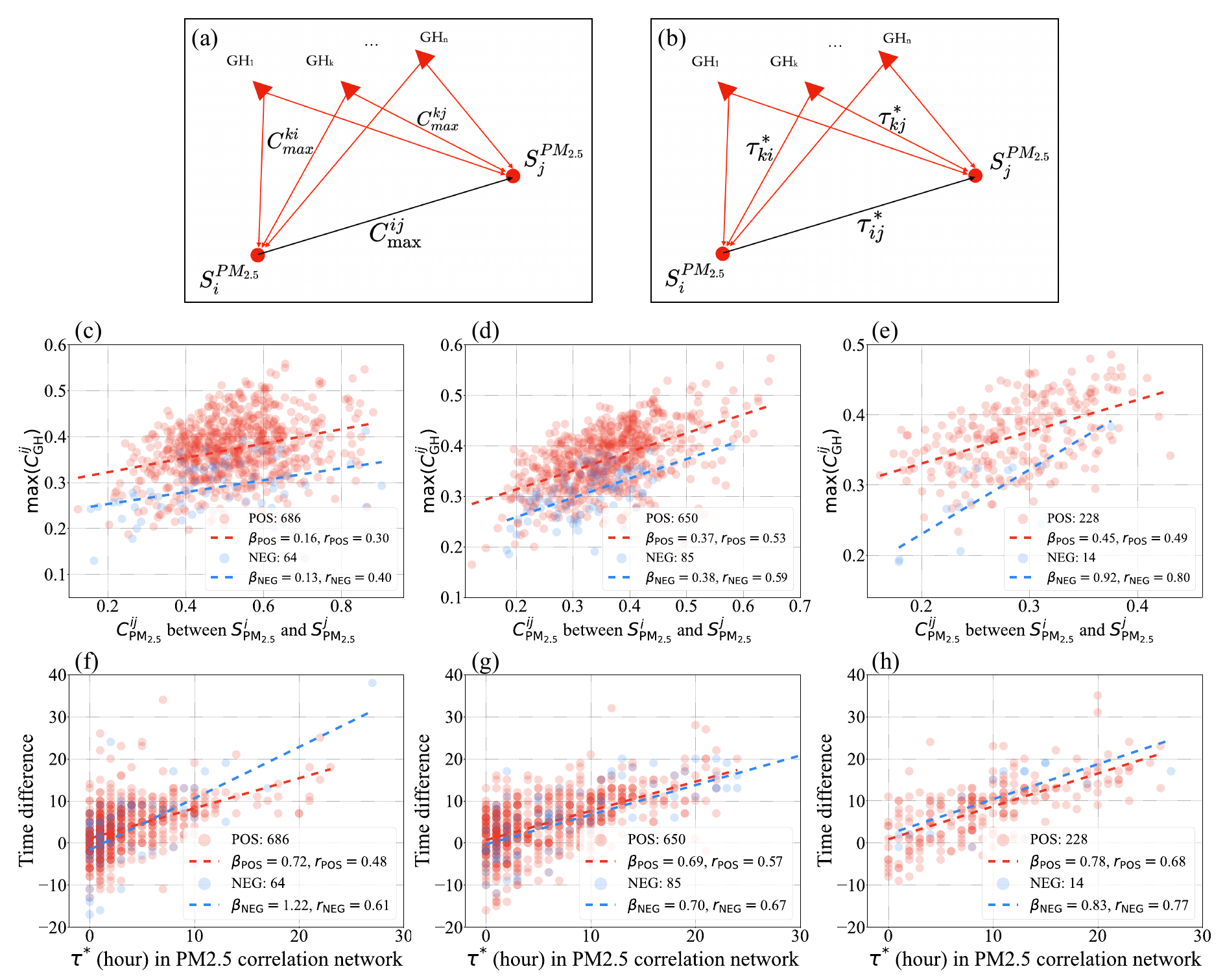}
  \caption{\textbf{Influence of geopotential height (GH) anomalies on long-range PM$_{2.5}$ links. (a–b)} Schematic diagrams illustrating two mechanisms by which GH anomalies influence pairs of PM$_{2.5}$ sites: (a) correlation strength ($C_{\max}$) and (b) time delay alignment ($\tau$). \textbf{(c-e)} Scatter plots based on 2015–2024 data showing the relationship between the strength of PM$_{2.5}$ links ($C_{\text{max}}^{ij}$) and the strongest shared geopotential height driver ($C_{\text{GH}}^{ij}$), stratified by link distance: (c) 0–500 km, (d) 500–1000 km, and (e) above 1000 km. \textbf{(f–h)} Corresponding analysis of the observed PM$_{2.5}$ time delay ($\tau_{ij}^*$) versus the time difference in GH–PM$_{2.5}$ time delays ($\tau_{kj}^* - \tau_{ki}^*$), using the GH site with the highest $C_{\text{GH}}^{ij}$. Each point represents one PM$_{2.5}$ link. Red and blue nodes denote links positively or negatively correlated with GH sites, respectively. Dashed lines indicate linear regression fits. Clear positive relationships are observed in both mechanisms, with slopes and correlation coefficients increasing with distance, indicating stronger geopotential height control on long-range PM$_{2.5}$ co-variability.}
  \label{fig.4}
\end{figure}

To further support our hypothesis regarding the influence of geopotential height anomalies on long-range PM$_{2.5}$ correlations, we conducted a systematic analysis of the temporal relationships among geopotential height (500hPa) sites and the correlated PM$_{2.5}$ grid sites (Fig. 4). Two conceptual frameworks are introduced: one focusing on correlation strength (Fig. 4a), and the other on time delay alignment (Fig. 4b). Specifically, we examine how either the strength of correlations between geopotential height sites and PM$_{2.5}$ sites, or the similarity in time delays from a common geopotential height site, may influence PM$_{2.5}$ link temporal characteristics. 

Fig. 4c–e correspond to the mechanism in Fig. 4a and examine the relationship between the correlation strength of a PM$_{2.5}$ link ($C_{\text {max}}^{ij}$) and the composite influence from geopotential height anomalies, quantified as the maximum $max(C_{\text{GH}}^{ij})$ among all GH sites jointly connected to both PM$_{2.5}$ endpoints. This composite measure captures the strongest common driver from geopotential height anomalies, where $C_{\text{GH}}^{ij} = \sqrt{C_{\text{max}}^{ki} \cdot C_{\text{max}}^{kj}}$, with $C_{\text{max}}^{ki}$ and $C_{\text{max}}^{kj}$ denoting the peak correlation strengths between GH site $k$ and PM$_{2.5}$ sites $i$ and $j$, respectively. Fig. 4f–h correspond to the mechanism in Fig. 4b and illustrate the relationship between the observed PM$_{2.5}$ time delay (${\tau_{i j}^*}$) and the difference in time delays between geopotential height sites and the two PM$_{2.5}$ sites ($\tau_{kj}^* - \tau_{ki}^*$). Since a single PM$_{2.5}$ link may be associated with multiple geopotential height sites, we select the one with the highest composite correlation strength $max(C_{\text{GH}}^{ij})$ to compute the corresponding time difference.

All results in Fig. 4 are derived from the full set of significant PM${_{2.5}}$ links identified over the ten-year period from 2015 to 2024. In both sets of analyses (Figs. 4(c–e) and 4(f–h)), PM$_{2.5}$ links are stratified by spatial distance: 0–500 km, 500–1000 km, and above 1000 km. Across both mechanisms, a consistent trend emerges: with increasing distance, the slope of the fitted regression lines becomes steeper, and the correlation coefficient ($r$) increases markedly. This progression suggests that while short-range PM$_{2.5}$ co-variability may still be influenced by local transport processes, long-range PM$_{2.5}$ links are more tightly coupled with large-scale geopotential height anomalies. These findings provide robust support for the hypothesis that mid-tropospheric circulation exerts a dominant influence on long-range air pollution synchronization in China.

\section{Conclusion}

In this study, we have systematically investigated the existence and the persistence of significant long-range correlations in PM$_{2.5}$ concentrations across China based on a ten-year high-resolution PM$_{2.5}$ dataset and advanced cross-correlation network analysis. Our findings reveal that PM$_{2.5}$ variations exhibit stable, repeated correlation structures, i.e., network links, not only at local scales but also at substantial distances exceeding 1000 km, a phenomenon that cannot be adequately explained by surface wind-driven transport alone due to the short observed delays. Further analysis of the 500 hPa geopotential height anomalies indicate a clear and significant meteorological control, where synoptic-scale atmospheric circulation patterns predominantly modulate these extensive pollution linkages. By categorizing significant PM$_{2.5}$ links based on their associations with geopotential height anomalies, we have illustrated that long-distance correlations are particularly influenced by large-scale atmospheric conditions rather than merely local emissions or surface transport mechanisms.

The progressively stronger alignment between geopotential height-driven delays and PM$_{2.5}$ co-variations with increasing spatial distances strongly confirms the critical role of upper atmospheric dynamics, suggesting that these large-scale meteorological factors should be explicitly considered in transboundary pollution assessments and air quality management policies. This enhanced understanding not only advances the scientific comprehension of pollution transport dynamics but also has practical implications for coordinated regional air-quality management strategies. The presence of stable and recurrent long-range links, shaped by large-scale circulation patterns, indicates a potential to forecast high-pollution events at distant locations several days in advance. Incorporating such network-based insights into existing early warning systems could improve the timeliness and spatial coverage of air quality alerts, supporting more proactive and spatially informed mitigation efforts. Future research should extend these findings by integrating additional atmospheric parameters and applying this methodological framework to other pollutant species and geographical contexts to further elucidate global-scale atmospheric pollution dynamics.

\acknowledgments
S.H. thanks the Israel Science Foundation (Grant No. 189/19 and 201/25), the NSF-BSF (Grant No. 2019740), the EU H2020 project DIT4TRAM,  the EU H2020 Project OMINO (Grant No. 101086321), the VATAT National Foundation  for Climate and the Israel Ministry of Innovation, Science \& Technology (grant number 01017980), for financial support.

\end{document}